# Long Term Planetary Habitability and the Carbonate-Silicate Cycle

Andrew J. Rushby[1,2], Martin Johnson[2,3], Benjamin J.W. Mills[4], Andrew J. Watson[5] and Mark W. Claire[6,7,8]

[1] NASA Ames Research Center, Moffett Field, CA
[2] School of Environmental Science, University of East Anglia, Norwich, UK
[3] Centre for Environment, Fisheries and Aquaculture Sciences, Pakefield Road, Lowestoft, UK
[4] School of Earth and Environment, University of Leeds, Leeds, UK
[5] College of Life and Environmental Sciences, University of Exeter, Exeter, UK.
[6] School of Earth and Environmental Sciences, University of St. Andrews, St. Andrews, UK
[7] Centre for Exoplanet Science, University of St. Andrews, St. Andrews, UK
[8] Blue Marble Space Institute of Science, 1001 4th Ave, Seattle, WA, USA

Corresponding author - Andrew Rushby email: andrew.j.rushby@nasa.gov, phone 650-604-6676

*Abstract*

The potential habitability of an exoplanet is traditionally assessed by determining whether or not its orbit falls within the circumstellar 'habitable zone' of its star, defined as the distance at which water could be liquid on the surface of a planet (Kopparapu *et al.*, 2013). Traditionally, these limits are determined by radiative-convective climate models, which are used to predict surface temperatures at user-specified levels of greenhouse gases. This approach ignores the vital question of the (bio)geochemical plausibility of the proposed chemical abundances. Carbon dioxide is the most important greenhouse gas in Earth's atmosphere in terms of regulating planetary temperature, with the long term concentration controlled by the balance between volcanic outgassing and the sequestration of $CO_2$ via chemical weathering and sedimentation, as modulated by ocean chemistry, circulation and biological (microbial) productivity. We develop a model incorporating key aspects of Earth's short and long-term biogeochemical carbon cycle to explore the potential changes in the $CO_2$ greenhouse due to variance in planet size and stellar insolation. We find that proposed changes in global topography, tectonics, and the hydrological cycle on larger planets results in proportionally greater surface temperatures for a given incident flux. For planets between 0.5 - 2 $R_\oplus$ the effect of these changes results in average global surface temperature deviations of up to 20 K, which suggests that these relationships must be considered in future studies of planetary habitability.

## *Introduction*

As the catalog of known exoplanets continues to grow, the potential habitability of planets discovered around other stars in our Galaxy remains of key interest to astrobiologists and planetary scientists. Detailed direct imaging of a given Earth-like planet and its atmosphere would allow us to assess its ability to host and support life, but contemporary exoplanet detection and characterization techniques are not yet sensitive enough to return the data required to make these inferences. Atmospheric spectroscopy can be used to obtain spectra from massive Jovian planets in close orbits around their stars ('Hot Jupiters') to determine atmospheric composition and circulation, as well as disequilibrium effects (e.g. Swain *et al*., 2008). Employing this technique for smaller planets is technically challenging, but has been achieved in certain cases where the transit depth (the ratio of planet to star size) is favourable, the first definitive example being spectra returned from three potentially Earth-like planets orbiting the ultracool dwarf TRAPPIST-1 (de Wit *et al*., 2016). Geophysical or geochemical modelling of exoplanetary systems is in its relative infancy as a discipline, but this approach is currently one of a very limited number of contemporary options for learning more about these enigmatic worlds.

Our knowledge of the geochemical cycles that have controlled atmospheric composition through the history of our home planet remains incomplete, but careful studies have elucidated what appear to be the controlling factors in the carbon cycle over the lifetime of the Earth (Berner, 1991; Berner, 2006; Sleep and Zahnle, 2001; Hayes and Waldbauer, 2006), and the resulting modelled changes in $CO_2$ concentration and global surface temperature appear reasonable when compared to what direct evidence is available (Royer et al., 2004; Mills et al., 2014). In this paper we adapt a model of the Earth's carbon cycle to infer the temperature and habitability for Earth-like planets of different sizes and star properties.

Earth appears to have maintained habitable (0-70 °C) conditions at its surface for as much as 4 Gyr in spite of significant increases in solar luminosity over this period (Rushby *et al.*, 2013; Sackmann *et al.*, 1993). This is due to the effect that increased solar forcing (or other non-$CO_2$ temperature forcing) has on the balance between the drawdown of atmospheric $CO_2$ via the weathering of silicate minerals, which is temperature dependent, and the tectonic and volcanic emission of $CO_2$ to the atmosphere, which operates independently of surface temperature. This silicate weathering feedback, first postulated by Walker, Hayes and Kasting (1981) has most likely led to a steady reduction in atmospheric $CO_2$ concentration over the lifetime of the Earth. Direct estimation of early Earth $CO_2$ concentration from geologic proxy records is difficult, but efforts to do so have often suggested upper limits (see review in Feulner (2012)), which generally do not provide the required radiative forcing to maintain liquid water at the planetary surface. By contrast, numerical models based on the carbonate-silicate cycle generally predict higher concentrations (e.g. $pCO_2$ ~100 PAL; Sleep and Zahnle, 2001) in line with the strong negative temperature feedbacks present – without liquid water the key $CO_2$ sink is removed and the atmospheric concentration can increase unbounded.

It should be briefly noted that determinations of radiative-forcing of large amounts of $CO_2$

remain under active debate (Halevy *et al*., 2009) and incorporations of newly measured laboratory parameters have continuously modified the "radiative effect" that a given amount of $CO_2$ will have. Differential incorporation of various physics (e.g. line absorptions at high temperatures, collisionally-induced absorption, pressure broadening) involved with radiative forcing of high $CO_2$ atmospheres have therefore allowed various re-computations of the 'circumstellar habitable zone' (e.g. Kopparapu *et al*., 2013; Kopparapu *et al*., 2016; Airapetian *et al*., 2016) in a manner that makes it difficult to directly compare studies, and their conclusions, over time.

Debate around the $CO_2$ greenhouse / surface temperature relation will continue, but it is generally accepted that on an Earth-like planet with active tectonics, gradual warming of the host star will cause atmospheric $CO_2$ to decline over time. In this paper, we investigate the significance of the carbonate-silicate weathering feedback in modifying habitable lifetimes relative to previously-developed simple energy balance approaches, focusing on predicting biogeochemically self-consistent $CO_2$ concentrations, and their evolution through planetary history. We extend the model to assess the first-order effect of planet size on this process, and determine how the drawdown of $CO_2$ over geological time effects the reorganization and eventual termination of life on Earth, and other potentially life-bearing planets of varying sizes.

*Methodology*

A carbon cycle model following Sleep & Zahnle (2001) (henceforth 'SZ01') is constructed to compute levels of atmospheric carbon dioxide ($pCO_2$) and globally-averaged surface temperature for a given incident flux in the range 0.75 to 1.25 $S_\oplus$, where 1 $S_\oplus$ corresponds to the present-day incident flux at the top of the atmosphere (1367 W m$^{-2}$). This model is constructed of four carbon reservoirs: Ocean/Atmosphere ($R_{oa}$), Oceanic Crust ($R_{oc}$),

Continental Sediment ($R_{cs}$), and the Mantle ($R_{man}$) (see Fig. 1). Here 'continental sediment' indicates both the carbonates and organics present in the continental crust as well as the C-bearing sediments on the continental shelves, which are geologically homologous (SZ01). Carbon is transferred through the system by a series of fluxes, constrained by the assumption of a steady state pre-industrial carbon cycle. The temporal resolution of the model is of the same order as the complete equilibration time of carbon in the ocean atmosphere system (~$10^6$ years).

Planetary surface temperature is computed from a 1-D radiative-convective climate model with an Earth-like atmospheric water vapour profile, derived from a 1-bar, cloud- and ozone-free, $N_2$-dominated model atmosphere, which includes $CO_2$ and $H_2O$ absorption coefficients from the HITRAN and HITEMP line-by-line databases (Haqq-Misra *et al.*, 2008; Kopparapu et al., 2013).

In order to determine $p$$CO_2$, carbon is separated between the ocean and the atmosphere by a partitioning constant ($\varphi$), which, assuming a pre-industrial $p$$CO_2$ of 280 ppm, is 0.016 (Bergman *et al.*, 2004). However, Omta et al. (2011) suggest an approximately exponential dependence between atmospheric $CO_2$ and average ocean temperature, so we include an additional scaling that alters the partitioning of $CO_2$ between the ocean and the atmosphere as a function of temperature. This allows for the model to account for large changes in ocean temperature, for example those that would occur towards the end of the planet's habitable lifetime as solar output increases steadily, on the partitioning of carbon species between the ocean and the atmosphere. Their scaling produces an approximate 4% increase in fractional atmospheric $CO_2$ per 1 K increase in ocean temperature, relative to present day ocean temperatures. To ensure mass balance, atmospheric volume was normalized to atmospheric scale height, which was computed as a function of $g$ under the assumption of an Earth-like dry air gas constant and mean atmospheric temperature of 288 K.

While planetary habitability is a complex and multifaceted concept that is not easily reducible to single parameter or metric, for the purposes of this work we consider habitability to be constrained to first order by average surface temperature and available $pCO_2$ for oxygenic photosynthesis by C3 and C4 terrestrial organisms (Cockell *et al*., 2016). Surface temperatures of between 273 and 343 K are applied as lower and upper bounds; 273 K representing a limit for 'Snowball' conditions and 343 K representing the upper limit, given that beyond this many of the proteins involved in photosynthesis begin to degrade (Rothschild, 2007). Excluding the survival of certain extremophile organisms on Earth which have been reported to endure temperature extremes ranging between 253 K and 395 K, these limits encapsulate the habitats of most life on the planet (Junge *et al*., 2004; Takai *et al*., 2008). However, setting an arbitrary habitable constraint at 273 K may be too limiting in terms of the stability of the climate, as a strong positive ice-albedo feedback will most likely result in an entirely glaciated planet up to a global average temperature of 278 K (Pierrehumbert, 2010). But conversely, it appears that the Earth may have strayed below this lower limit during two distinct periods around 2.3 Ga and 700 Ma (Kirschivink et al., 2000; Hoffman et al., 1998), not necessarily to the detriment of the evolving biosphere.

Photosynthetic primary producers on Earth require atmospheric carbon dioxide to produce organic compounds, producing oxygen as a by-product of this process. Low levels of atmospheric $CO_2$ will inhibit these organisms' metabolic function, and will therefore also affect levels of atmospheric oxygen crucial for the metabolisms of most eukaryotic organisms. The vast majority of terrestrial carbon capture is performed by plants utilising either the C3 or C4 photosynthetic pathway (representing 89% of plant species, combined) (Bear & Rintoul, 2016). The effective $pCO_2$ limit for these organisms is ~150 and ~10 ppmv, respectively (Lovelock & Whitfield, 1982; Caldeira & Kasting, 1992; O'Malley-James *et al*.., 2013; O'Malley-James *et al*.., 2014). Below these limits, photosynthesizers that use these carbon concentration mechanisms become less able to effectively capture carbon from the atmosphere in order to build biomass. Given the importance of these organisms as primary producers in the terrestrial biosphere, we conclude that the biosphere would effectively undergo a significant reorganization following their extinction, possibly resembling that of the anoxic, early history of Earth .

*i. Planet Mass-Radius Relationships*

An understanding of the interiors of exoplanets requires (at a minimum) knowledge of their mass and density. Radial velocity techniques allow for planet mass to be estimated (up to a degeneracy in the orbital inclination) based on the gravitational effect that the orbiting world has on its host star. Using transit photometry, the radius of an orbiting planet can be estimated based on the amount of light that the planet obscures, and its mass can be determined if the composition of the interior of the planet is known (Jenkins *et al*., 2002). However, radius and mass estimates are not sufficient to identify a planet's interior composition, with the possible exception being at the extremes in density – very low density planets are likely to be dominated by H and He, and very high density planets are likely composed primarily of Fe (Sotin *et al*., 2010). Scaling relationships between planet radius and mass can be used to constrain geodynamic processes operating on exoplanets to first order, if a series of simplifying assumptions are made regarding bulk composition, surface gravity, geothermal heat flux, as well as mantle and core radii (Sotin *et al*., 2010; Seager *et al*., 2007). As mass is a product of the volume of the planet and its density, it scales with the approximate cube of the radius, modified by the effect of the bulk composition of the interior. Although some uncertainty exists as to the upper radius limit for rocky worlds (e.g. Rogers, 2015), we choose an optimistic upper limit of 2 $R_\oplus$ and a lower limit of 0.5 $R_\oplus$. We then adopt the mass-radius relationships presented in Barnes *et al.* (2015):

$$M/M_\oplus = (R/R_\oplus)^{3.268} \quad (\leq 1 \ R_\oplus)$$

$$M/M_\oplus = (R/R_\oplus)^{3.65} \quad (> 1 \ R_\oplus)$$

Eq.1

By scaling the normalized size of the crustal reservoirs of carbon using these relationships, the effect of increasing mass on carbon storage and transfer between these reservoirs can be quantified.

*ii. Volcanism*

Some mode of volcanic outgassing that enables atmospheric volatile replenishment is considered to be a necessary prerequisite for long-term planetary habitability due to the ability of volcanic outgassing to significantly alter the composition of planetary atmospheres (Korenaga, 2012; Kasting & Catling, 2003). However, it is currently unclear as to whether the tectonic regime observed on the Earth is typical of terrestrial planets of similar size, or indeed how these processes will change with planet size (Kite et al., 2009). Volcanism is expected to be a widespread characteristic of rocky planets resulting from partial melting of their silicate mantle under extremes of pressure, but contention exists over the form of thermal convection that would dominate on massive planets, and to what alternative form their tectonic regimes may take (Kite *et al.*, 2009; O'Neill & Lenardic, 2007; Valencia *et al.*, 2007).

Regardless of the mechanism of convection, internally-generated heat must escape, following the laws of thermodynamics. Assuming that the abundance of radiogenic elements in a planet is proportional to its mass, then the radiogenic heat flux (Q) varies as a function of planet mass, according to Sotin et al. (2010):

$$Q = Q_\oplus (M/M_\oplus)^{0.452}$$

Eq.2

Where $Q_\oplus$ is the geothermal heat flux on the Earth. While acknowledging considerable uncertainty regarding the rate of geothermal heat loss during the Archean, this model follows the precedent of SZ01 (as well as Lowell & Keller, 2003; Hayes & Waldbauer, 2006; Franck & Bounama, 1999) in assuming a time-dependent decay function in the form:

$$Q_\oplus = 1 + (t/t_0)^{-0.7}$$

Eq.3

where $t_0$ describes the time of planet formation. This parameterization predicts heat loss rates from the early Earth between 4.5 and 3 times present day. This relationship has a significant control on the likely geochemical and climatic evolution of the planet as it is a key driver in many of the flux terms associated with the carbon cycle. In geodynamic models, the seafloor spreading rate is found to closely approximate $Q^2$, and has a strong control over the rate of subduction of oceanic crust into the mantle, and by implication the amount of carbon returned to the ocean/atmosphere system via arc volcanism ($F_{subarc}$) as well as via ridge degassing ($F_{ridge}$), metamorphism ($F_{meta}$), and the rate of seafloor hydrothermal carbonatization ($F_{hydro}$) (Franck et al., 1999; SZ01), which are defined in Table 1 and Fig.1)

### *iii. Weathering*

Silicate weathering and its dependence on temperature and atmospheric $CO_2$ is an integral driver controlling long-term planetary climate. A negative feedback between the rate of weathering of silicate rocks, the surface temperature of the planet and the partial pressure of carbon dioxide in the atmosphere operates to regulate $pCO_2$ over geological time ($>10^6$ years): as surface temperatures increase (i.e. due to steadily increasing solar luminosity), rates of continental weathering increase, thereby drawing down $CO_2$ and reducing the overall radiative forcing (Walker et al., 1981). Studies find the rate of $CO_2$ sequestration via continental silicate weathering is a strong function of temperature and a weak function of $pCO_2$. Our formulation of this process, following SZ01, is:

$$F_{SiO_3}w = F_{SiO_3}w_0 \cdot pCO_2^{\beta} \cdot exp\left(\frac{T - T_0}{B}\right) \cdot \kappa \cdot R_E$$

Eq.4

here $\beta$ is an exponent that controls the sensitivity of the reaction to the partial pressure of atmospheric carbon dioxide and the rate of carbonic acid formation (held at 0.3 for this work (SZ01)) and $B$ is a dimensionless weathering calibration parameter suggested by Walker et al., (1981) to account for runoff ($B = 13.7$). The additional parameter $\kappa$ simulates the effects of biotic enhancement of weathering, where terrestrial organisms, via the action of their roots, metabolisms or exuded organic acids, increase the rate of weathering of silicate minerals beyond what would be expected in an abiotic environment (Berner, 1997; SZ01; Schwartzman & Volk, 1989). The delivery of bicarbonate ions and silicate cations to the ocean is thought to be strongly dependent on the supply of material via erosion, and by precipitation and runoff (West et al., 2005; Hartmann et al., 2014). The additional scaling parameter $R_E$ in Eq. 4 represents our first-order expectation that evaporation rates will scale with the fraction of planetary surface area covered by ocean. $R_E$ captures both variable ocean/continental fractions and associated carbon partitioning between oceanic and terrestrial environments, as well as evaporation rates from those surfaces. A variable ocean/continental fraction is given in this zero-D formulation as a parameter that determines the contrast between the albedo of the ocean and the land surface, which in turn affects the energy balance at the surface. The initial carbon inventory of the ocean and continental reservoirs is then proportionally partitioned relative to the steady-state, present-day Earth case.

A comparison with the COPSE geochemical model (Bergman *et al.*, 2004) was carried out under conditions of differing biotic weathering enhancement (κ), and our modelled $pCO_2$ output demonstrated good agreement with that model over the Phanerozoic (Fig 2a). COPSE has a greater temporal resolution, and also includes a sulfur (S), phosphorus (P) and partial iron (Fe) cycle, important for accurately representing the global oxygen cycle, which produces more detailed, high resolution *p*CO$_2$ data, as output is constrained by comparison to $\delta^{34}S$ and $\delta^{13}C$ isotope records. As it is clear from this comparison that the model developed here lacks the ability to resolve highly temporally variable $CO_2$ and temperatures associated with the Phanerozoic, mean values of $CO_2$ and temperature from COPSE (split at the Permo-Carboniferous) have also been included for a more appropriate comparison. Based on this comparison, we hold *κ* at 0.5 for this model, representing a doubling of terrestrial weathering rates by biology. Furthermore, our model output demonstrates reasonable agreement with available geochemical proxy data from Feulner (2012). Fig.2b (modified from Feulner (2012)) displays *p*CO$_2$ as function of S/S$_\oplus$ generated from default model conditions, as well as estimates of atmospheric carbon dioxide from other sources. Disparity between this work at that of Rosing (2010) can be explained by their underestimate of surface albedo.

*Results*

When coupled with a stellar evolution model that predicts luminosity over time, our model allows for estimation of the duration of habitable conditions on Earth-like worlds. For example, using the stellar evolution data from the model presented in Rushby *et al*., (2013) and setting an upper temperature limit of 343 K, the habitable period of an Earth-like world around the Sun can be quantified (Fig. 3a). For a 1 $R_\oplus$ planet at 1 AU (e.g. Earth), this limit is approximately 6.23 billion years (Gyr) after planet formation, or 1.69 Gyr from present day. Additionally, atmospheric $CO_2$ falls below the C3 and C4 plant compensation limit after ~5.37 Gyr and ~5.89 Gyr respectively, which would put increasing stress on primary producers dependent on oxygenic photosynthesis for energy production (Fig. 3b). The carbonate-silicate buffer becomes ineffective at this stage, as $CO_2$ is not returned to the atmosphere/ocean system at an equivalent rate as it is being removed due to the lower geothermal heat flux, ocean-floor spreading and tectonic outgassing rates.

Notably, these 'photosystem' compensation limits are reached some time before the model predicts temperatures to be above the habitable maximum, suggesting that considering 'habitability' merely as a product of incident flux and surface temperature is too limited and neglects the inherent complexities of the interconnected planetary system. That considered, regardless of planet size (between 0.5 and ~2 $R_\oplus$) a 'runaway greenhouse' event (set at the limit when outgoing longwave radiation (OLR) is greater than incoming shortwave radiation) occurs in our model when incident flux on the planet is between ~1.13 and 1.15 $S_\oplus$. We note that there is a very slight planet size dependence on the timing of the 'moist greenhouse' transition that occurs just prior to full runaway (beginning around ~1.1 $S_\oplus$). However, the radiative effect of these slight differences is negligible as $CO_2$ remains uniformly low across the range of planet sizes at this stage because the extreme enhancement in silicate weathering at very high insolation overwhelms minor differences in $CO_2$ content and outgassing rates. Larger planets are, however, warmer during the early stages of the simulation when incident flux is lowest and $CO_2$ concentration highest, but given the limitations of this model in terms of representing the likely surface and atmospheric conditions of the early history of a terrestrial planet (for example, the hydrogen/helium paleoatmosphere, a 'magma ocean' phase, or period of heavy bombardment) this result is likely not fully representative of the actual conditions we may expect during this period. Our primary focus is the stage at which high stellar flux and secularly decreasing $p$CO$_2$ begins to dismantle the carbonate-silicate buffer, which in the Sol-Earth system is towards the end of the Sun's main-sequence lifetime, but several billion years before its red giant phase.

This model predicts that planets with larger radii tend to have higher average surface temperatures for a given incident flux during their habitable lifetimes (Fig 3a and Fig 4). For example, a 2 R$_\oplus$ planet receiving the same incident flux as the present Earth is ~5 K warmer on average than an Earth-sized world (Fig. 4). The reasons for this are manifold: higher geothermal fluxes (that also persist for longer after the planet's formation due to a more significant radiogenic inventory) leading to greater rates of outgassing of CO$_2$, subduction and seafloor spreading contribute to higher average $p$CO$_2$ values for larger planets (Fig.3b, Fig.5). Terrestrial weathering fluxes to the oceans are also more significant due to greater chemical weathering and erosion of continental materials, somewhat but not entirely counteracting the increased outgassing and leading to more rapid cycling of carbon through the planetary system overall.

Atmospheric CO$_2$ levels on larger planets remain higher for a given incident flux despite this enhanced terrestrial weathering. They also remain above the carbon fixation limits of photosynthesizing primary producers for proportionately longer, which would in turn allow for a biosphere based on oxygenic photosynthesizers to persist for longer than on smaller planets. As an example, this model suggests that a 2 R$_\oplus$ planet at 1 AU around the Sun would maintain $p$CO$_2$ levels above the C3 and C4 photosystem limits for ~5.5 Gyr and ~5.93 Gyr, respectively. This is 130 and 40 million years longer than may be expected on an Earth-sized planet at the same separation. As the 'runaway greenhouse' limit discussed above seems to be mass-independent (over the range of planetary masses considered in this work), the primary control that planet size exerts on long-term habitability is through the recycling of carbon through the planet system over geological time.

Conversely, planets smaller than 1 R$_\oplus$, due to their smaller radiogenic inventories and associated heat flux, experience less vigorous rates of outgassing, subduction, seafloor spreading and terrestrial weathering. A 0.5 R$_\oplus$ planet is approximately 2 K cooler than a 1 R$_\oplus$ planet at a radiative flux of 1 S$_\oplus$. At lower incident flux, this effect is especially significant: at 0.75 S$_\oplus$ (approximately equivalent to the flux incident on the Earth during the early history of the Solar System) the temperature difference between a 0.5 R$_\oplus$ and 1 R$_\oplus$ planet is closer to 8 K, and nearly 20 K between 0.5 R$_\oplus$ and 2 R$_\oplus$. To further illustrate this, if Mars was ocean-covered at Earth's current position it would be in a Snowball state, even if the planet was allowed all biogeochemical cycling process available to Earth. This is partly due to the fact that the atmospheres of these smaller worlds are also proportionately depleted in carbon, which is removed at a greater rate (assuming burial continues up to the limit of the extinction of a C3 and C4 biosphere) due to the earlier cessation of geophysical activity that prevents the return of previously subducted carbon to the atmosphere. While not captured in this particular model, this process would be further exacerbated by the fact that atmospheres of smaller worlds are more rapidly stripped away by high energy stellar particles. C3 plants on an Earth of half the radius, but at the same orbital separation, would become stressed by lack of atmospheric carbon ~5.1 Gyr after planet formation. In terms of atmospheric carbon availability for photosynthesis habitability, planet size matters.

Applying this model to known exoplanets requires the assumption that these worlds will resemble Earth in many of their physical and chemical characteristics. At this stage, this assumption remains defensible, given the limited information available to astrobiologists on the interiors and atmospheres of small, potentially rocky worlds. Allowing this first-order assumption allows for re-evaluation of planetary habitability in some cases. For example, also plotted in Fig. 4 (red marker) alongside the Earth is GJ 667Cc: a possibly terrestrial planet ($R_\oplus = 1.54$, $S_\oplus = 0.88$) discovered orbiting a nearby red dwarf star (Bonfils *et al.*, 2011). The effective temperature for this planet, assuming an albedo of 0.3, is in the range of 227 K - 247 K, depending on uncertainties in its orbital semi-major axis and stellar parameters. Adopting the flux and size estimates given above, and assuming an Earth-like carbonate-silicate cycle, ocean/continent fraction and $CO_2$ greenhouse, GJ 667Cc could maintain a significant, biogeochemicaly self-consistent $CO_2$/$H_2O$ greenhouse (assuming the same relative initial conditions as the Earth), with a predicted surface temperature buffered closer to 283 K. This value is significantly warmer than previous estimates, rendering this world potentially more 'habitable' than previously considered, if the various "Earth-like" assumptions hold.

This case study can be taken a step further, should the relative age of a planet's star be known. GJ 667Cc orbits a small, red dwarf star ($M_\odot = 0.31$) that has an estimated main-sequence lifetime of 288 Gyr (Rushby *et al.*, 2013), well in excess of the current age of the Universe. The implications of this long main-sequence evolution is the possibility that GJ 667 Cc may have a habitable period on the order of 65 Gyr, excluding the potentially deleterious effects of the frequent flaring to which these small stars are prone, as well as the effects of tidal locking and tidal heating, which may be considerable and increasingly important as internal heat sources diminish over time (Shields *et al.*, 2016). However, the age of its host star is poorly constrained at >2 Gyr, and therefore any estimate of the current duration of habitable conditions on GJ 667Cc is subject to considerable uncertainty.

*Discussion*

This work presents a simple biogeochemical model that seeks to represent the evolution of the biosphere-planet system over geological and astronomical time, with particular focus on the carbonate-silicate cycle due to its significant effect on the surface, interior and atmosphere of the planet. Larger planets tend, on average, to be warmer than smaller planets under a given stellar flux, with this effect especially pronounced at lower insolation when the atmospheric concentration and radiative effect of $CO_2$ is more significant. Using temperature and $CO_2$ concentrations as a first approximation of habitable conditions, it seems that the complex interplay between these components produces an uncertain conclusion: habitability is strongly variable in time, and also falls along a spectrum as opposed to a clear binary distinction between 'uninhabitable' and 'habitable'. Conditions exist where, whilst average surface temperatures may fall within the traditional habitable boundaries, other factors act to reduce the overall habitability of the planet.

The upper temperature boundary considered here represents the endmember case for habitability, but before this limit is reached the terrestrial biosphere will undergo significant and likely dramatic reorganisation, in terms of spatiotemporal distribution, habitat and resource availability, and metabolic stress exacerbated by falling $pCO_2$ levels, which results in a form of 'photosystem-limited habitability' set by this $CO_2$ compensation limit. The planet enters this state towards the end of its habitable lifetime as surface temperatures increase under a brightening star, and the action of terrestrial weathering draws down $CO_2$ at a rate faster than it can be returned to the atmosphere by volcanic sources, the driver of which (geothermal heat flux) is also diminishing with time as the planet's radiogenic inventory becomes depleted. Primary producers crucial for supporting the planet's biosphere would become $CO_2$-stressed, and this would eventually result in the collapse of the (terrestrial) biosphere long before the planet becomes too warm. O'Malley-James et al. (2014) describe the changes likely to occur in the trophic hierarchy of the terrestrial biosphere of a planet in a state of $CO_2$-stress: below 150 ppm $CO_2$, desiccation tolerant CAM and C4 plants would briefly dominate in isolated refugia (subterranean caves, high altitude environments), possibly followed by a period of microbial anoxygenic photosynthesis when $CO_2$ falls below 10 ppm. Below 1 ppm $CO_2$, the end of all photosynthetic life is expected, and this juncture would represent a terminal extinction event for the biosphere, even if surface temperatures remained within habitable limits (O'Malley-James et al., 2014). It should be noted that we assume a 'business as usual' scenario and that, given the timescales over which this $CO_2$ decline takes place, it may be feasible, but outside of the scope of this paper, to suggest that evolutionary adaptations may take place to extend the lifespan of a terrestrial photosynthetic biosphere to a low-$CO_2$ world.

Furthermore, this work has demonstrated that a planet's size affects the maximum limit of habitable conditions through geophysical and atmospheric processes, one that can be, to some degree of accuracy, derived from the mechanics of carbon cycle as we currently understand it. Larger terrestrial planets are predicted to have warmer average surface temperatures and more carbon in their atmospheres through the control that planet mass has on various elements of the carbonate-silicate cycle. Planetary mass-dependent effects include a greater and more persistent geothermal flux that drives subduction and seafloor spreading at an accelerated rate, and for longer, on more massive terrestrial planets. Furthermore, the products of enhanced terrestrial weathering are delivered to the oceans faster due to greater erosion of continental materials on more massive planets. This results in higher average $p$CO$_2$ values for larger planets, which in turn generates a more effective greenhouse. However, the mass of a planet seems to have little effect on the eventual initiation of a runaway greenhouse event in this case; models of atmospheric photochemistry that take into consideration surface pressure and pressure broadening on larger worlds would likely suggest otherwise, and would represent a potential future improvement to this particular model.

An implication of these results is that we are ~70% of the way through the habitable lifetime of the Earth (4.54 Gyr of 6.23 Gyr). We can begin to contrast and compare this value with that of known exoplanets, if the age of their star is known and by adopting the assumptions outlined above, to determine the comparative temporal habitability of worlds across the Galaxy. As demonstrated by the case of GJ 667Cc (a potentially rocky 1.54 R$_\oplus$ planet in the orbit of a red dwarf binary) planets in the orbit of small stars with long main-sequence lifetimes may considerably eclipse Earth in this measure. Despite the fact that this concept is often overlooked in studies of planetary habitability, the duration and relative timing of habitable conditions on terrestrial planets also has pressing implications for atmospheric evolution and the future studies concerned with the spectroscopic detection of biosignature gases (Caldeira & Kasting, 1992; O'Malley-James et al., 2013; O`Malley-James et al., 2015; Rushby *et al*., 2013, Watson, 2008).

Furthermore, the nature of main sequence stellar evolution (i.e growing in luminosity over time) also presents the possibility of a once cold planet that formed beyond the outer edge of the habitable zone 'entering' the HZ during the course of its host star's main sequence lifetime. Rushby *et al*., (2013) suggests this may occur for Mars in our Solar System, but that it is likely too small and geologically inactive for habitable conditions to arise during the late stages of the Sun's main sequence lifetime. The possibility of the reactivation of a carbonate-silicate cycle on a planet undergoing secular stellar warming has not been studied in detail, but given that active plate tectonics is required to drive the subduction and outgassing of carbon from the planet's interior, it may not be possible to invoke this as a possibility. To further obfuscate the possibility of a late-stage reactivation of planetary carbonate-silicate cycling, the importance of mantle hydration as a mechanism for facilitating and maintaining plate tectonics over geological time remains unclear (Korenaga, 2010; Korenaga, 2012). It is possible that a drier mantle and lower associated viscosity will make the initiation of plate tectonics easier, but that maintaining a continental crust may require interaction between the hydrosphere and solid Earth over geological time (*ibid*).

Attempting to quantify the likely lifetime of the biosphere of habitable planets has been a key area of research in astrobiology and the concept remains extremely pertinent to astrobiology and SETI campaigns. Arguably, providing an estimate for the maximum lifespan of a planet's biosphere goes beyond a diagnostic investigation of the operation of feedback processes on the planet. This limit provides a terminal limit for biological evolution and speciation, with some bearing on certain parameters contained within the Drake Equation, such as the fraction of life-bearing planets on which intelligent life emerges, as well as the the duration of the civilisations formed by that life. Watson (2008), drawing on the earlier work of Carter (1983) and Szathmary & Maynard-Smith (1995), developed a probabilistic model that attempts to quantify the timing and likelihood of several critical (defined broadly as only occurring once in Earth history) steps in evolution en route to intelligent observer species. A key parameter in these calculations is the habitable period of the planet, which exhibits a strong control over the expected timing on these critical steps. Their broad conclusions are that only a planet with a long habitable period ($>10^9$ years) could support complex life, and intelligent life may require longer still, as these critical steps are incredibly unlikely - on the order of $10^{-9}$ $yr^{-1}$.

(Watson, 2008). Therefore, both future SETI studies as well as those searching for 'non-intelligent' life should prioritize exoplanets that satisfy both criteria of spatial and temporal habitability.

The idealised planetary system depicted here represents a first approximation of the geochemical cycling of carbon on a terrestrial world with a differentiated oceanic and continental crust. Assumptions implicit in the operation of the carbonate-silicate cycle are a limitation to this work however; it is not known at this stage if analogous processes will operate on terrestrial exoplanets, even those with oceans and active tectonics, nor how planetary composition and density will scale with size and what effect the formation history of the system will have on the eventual composition of the planetary body. Furthermore, our understanding of the operation of this cycle on the Earth reveals potential biological amplification and mediation at several junctures, including the fixation and burial of inorganic carbon in the ocean and the acceleration of mechanical erosion on the surface by the action of roots and exuded organic acids (Lenton & Watson, 2004; Mills et al., 2014). Additionally, plants return a large proportion of carbon-rich organic matter for burial upon their deaths, and throughout their lifecycle represent a significant interface between the geosphere and biosphere. Furthermore, the marine ecosystem supported by cyanobacterial photoautotrophs will likely persist for some time following the extinction of the terrestrial biosphere. The carbon capture and concentration mechanisms of these organisms, as well as the behaviour of $CO_{2(aq)}$ ensures that the oceanic carbon cycle will continue to operate, albeit under conditions of increasing stress and ecosystem reorganization. This is an area of study into which future iterations of this model will attempt to make ground. Separating biotic and abiotic processes implicit in the carbonate-silicate cycle and understanding their relative influence of the evolution of atmospheric $CO_2$ therefore remains a research priority in Earth system and exoplanetary system science alike.

**Conclusion**

This study is arguably one of the first to attempt to incorporate a self-consistent carbon cycle and radiative-convective climate model to determine how planet size affects the long-term cycling of carbon, as well as how these differences may alter the surface temperature of potentially habitable terrestrial planets. Here we demonstrate that both average planetary surface temperature and atmospheric carbon dioxide increase with increasing radius, and that we might expect a larger planet to be proportionately warmer and $CO_2$-rich for a given incident flux. This model can also be used, in conjunction with stellar evolution data, to refine estimates the duration of habitable conditions on a planet with similar carbonate-silicate cycling and compositional and geophysical characteristics to Earth, which is predicted to remain within habitable temperature bounds for approximately a further 1.7 billion years. However, the abundance of $pCO_2$ will decline secularly towards this limit, potentially starving terrestrial photosynthesizers of a suitable source of energy, and limiting planetary habitability through their inability to support a complex eukaryote biosphere as primary producers.

**Acknowledgements**

AJR would like to recognise the support of a Dean's Fellowship at the University of East Anglia, under which most of this work was produced, as well as an appointment to the NASA Postdoctoral Program at NASA Ames Research Center, administered by Universities Space Research Association under contract with NASA, where the model was updated and paper compiled. BJWM acknowledges a University of Leeds Academic Fellowship. AJR would like to thank Christian Clanton and Jessica Lewis for their useful notes on the early manuscript, as well as the invaluable feedback from two anonymous reviewers that strengthened the final version.

**Conflict of Interest**

The authors report no conflicts of interest.

*References*

Airapetian, V.S., Glocer, A., Gronoff, E., Hébrard & Danchi, W. (2016). Prebiotic Chemistry and atmospheric warming of early Earth by an active young Sun. *Nature Geoscience* **9** doi:10.1038/ngeo2719

Barnes, R., Meadows, V., & Evans, N. (2015). Comparative Habitability of Transiting Exoplanets. *The Astrophysical Journal*. **814** pp.91 – 102

Bear, R., Rintoul, D., Synder, B., Smith-Caldas, M., Herren, C., & Horne, E. (2016). *Principles of Biology* Open Access Textbooks, **1**.

Bergman, N.M., Lenton, T.M. & Watson, A.J. (2004). *American Journal of Science*, *304*, 397 - 437

Berner, R.A. (1991). A model for atmospheric CO2 over Phanerozoic time. *American Journal of Science* **291** pp. 339 - 376

Berner, R.A. (2006). GEOCARBSULF: A combined model for Phanerozoic atmospheric O2 and CO2. *Geochemica et Cosmochemica Acta* **70**(23) pp. 5653 - 5664

Berner, R.A (1997). The rise of plants and their effect on weathering atmospheric $CO_2$. *Science* **276** pp.544 - 546

Bonfils, X., Delfosse, X., Udry, S., Forveille, T., Mayor, M., Perrier, C., Bouchy, F., Gillon, M., Lovis, C., Pepe, F., Queloz, D., Santos, N.C., Segransan, D., & Bertaux, J., -L. (2011). The HARPS search for southern extra-solar planets. XXXI. The M-dwarf sample. *Astronomy and Astrophysics* **549** A109

Caldeira, K. & Kasting, J. F. (1992). The life span of the biosphere revisited. *Nature* **360** pp. 721 – 723

Carter, B. (1983). The anthropic principal and its implications for biological evolution.

*Philosophical Transactions of the Royal Society of London A*. **310** pp. 347 - 363

Cockell, C.S., Bush, T., Bryce, C., Direito, S., Fox-Powell, M., Harrison, J.P., Lammer, H., Landenmark, H., Martin-Torres, J., Nicholson, N., Noack, L., O'Malley-James, J., Payler, S., Rushby, A. J., Samuels, T., Schwendner, P., Wadsworth, J., & Zorzano, M.P. (2016). Habitability: A Review. *Astrobiology* **16**(1) pp. 89 - 117

Davies, J.H. & Davies, D.R. (2010). Earth's surface heat flux. *Solid Earth* **1** pp. 5 - 24

de Wit, J., Wakeford, H. R., Gillon, M., Lewis, N. K., Valenti, J. A., Demory, B.-O., Burgasser, A. J., Burdanov, A., Deirez, L., Jehin, E., Lederer, S. M., Queloz, D., Triaud, A. H. K. J., & Van Grootel, V. (2016), *Nature*, *537*, 7618, 69 -72

Driese, S. G., M. A. Jirsa, M. Ren, S. L. Brantley, N. D. Sheldon, D. Parker, & M. Schmitz (2011), Neoarchean paleoweathering of tonalite and metabasalt: Implications for reconstructions of 2.69 Ga early terrestrial ecosystems and paleoatmospheric chemistry, *Precambrian Research* **189**(1–2), 1–17, doi:10.1016/j.precamres.2011.04.003.

Franck, S. & Bounama, C. (1999). Continental growth and volatile exchange during Earth's evolution. *Physics of the Earth and Planetary Interiors* **100** pp. 189 - 196

Franck, S., Kossacki, K. & Bounama, C. (1999). Modelling the global carbon cycle for the past and future evolution of the earth system. *Chemical Geology* **159** pp. 305 - 317

Feulner, G. (2012). The Faint Young Sun Problem. Rev. Geophys. **50** RG2006. doi:10.1029/2011RG000375

Halevy, I., R. T. Pierrehumbert, and D. P. Schrag (2009). Radiative transfer in $CO_2$-rich paleoatmospheres. *Journal of Geophysical Research* **114** D18112, doi:10.1029/2009JD011915.

Haqq-Misra, J. D., Domagal-Goldman, S. D., Kasting, P. J., & Kasting, J.F. (2008). A Revised, Hazy Methane Greenhouse for the Archean Earth. *Astrobiology* **8**(6) pp.1127 - 1137

Hartmann A., Gleeson T., Rosolem R., Pianosi F. & Wagener T. (2014). Understanding recharge elasticity through large-scale simulations of Europe's karst regions under varying climatic boundary conditions. *EGU General Assembly* **16**, EGU2014-7976.

Hayes, J.M. & Waldbauer, J.R. (2006). The carbon cycle and associated redox processes through time. *Philosophical Transactions of the Royal Society B* **361** pp.931 – 950

Hessler, A. M., D. R. Lowe, R. L. Jones, & D. K. Bird (2004), A lower limit for atmospheric carbon dioxide levels 3.2 billion years ago. Nature 428, 736–738, doi:10.1038/nature02471.

Hoffman, P.F., Kaufman, A.J., Halverson, G.P. & Schrag, D.P. (1998). A Neoproterozoic

Snowball Earth. *Science* **281**(5381) pp.1342-1346

Jenkins, J. M., Caldwell, D. A., Borucki, W. J. (2002). Some Tests to Establish Confidence in Planets Discovered by Transit Photometry. *The Astrophysical Journal* **564** pp. 495 - 507

Junge, K., Eicken, H., & Deming, J.W. (2004). Bacterial Activity at −2 to −20°C in Arctic Wintertime Sea Ice. *Applied Environmental Microbiology* **70** pp. 550 - 557.

Kah, L. C., & R. Riding (2007), Mesoproterozoic carbon dioxide levels inferred from calcified cyanobacteria, Geology 35, 799–802, doi:10.1130/2FG23680A.1.

Kasting, J. F. & Catling, D. (2003). Evolution of a Habitable Planet. *Annual Reviews in Astronomy and Astrophysics*. **41** pp.429 – 463

Kaufman, A. J., & S. Xiao (2003), High CO2 levels in the Proterozoic atmosphere estimated from analyses of individual microfossils, Nature 425, 279–282, doi:10.1038/nature01902.

Kirschvink. J.L., Gaidos, E.J., Bertani, E.L., Beukes, N.J., Gutzmer, J., Maepa, L.N. & Steinberger, R.E. (2000). Paleoproterozoic snowball Earth: Extreme climatic and geochemical global change and its biological consequences. *PNAS* **97**(4) pp.1400-1405

Kite, E.S., Manga, M. And Gaidos, E. (2009). Geodynamics and rate of volcanism on massive Earth-like planets. *The Astrophysical Journal* **700** pp. 1732 - 1749

Kopparapu, R. K., Ramirez, R., Kasting, J. F., Eymet, V., Robinson, T. D., Mahadevan, S., & Deshpande, R. (2013), *ApJ*, *765*, 2, 131.

Kopparapu, R.K., Wolf, E.T., Haqq-Misra, J., Yang, J., Kasting, J.F., Meadows, V., Terrien, R., & Mahadevan, S. (2016). The Inner Edge of the Habitable Zone For Synchronously Rotating Planets Around Low-Mass Stars Using General Circulation Models. *The Astrophysical Journal* **819** (1) doi:10.3847/0004-637X/819/1/84

Korenaga, J. (2010). On The Likelihood Of Plate Tectonics On Super-earths: Does Size Matter? *The Astrophysical Journal: Letters* **725** L43

Korenaga, J. (2012) Plate tectonics and planetary habitability: current status and future challenges. *Annals of the New York Academy of Sciences* **1260** pp. 87 - 94

Lenton, T. M., & Watson, A.J. (2004). Biotic enhancement of weathering, atmospheric oxygen and carbon dioxide in the Neoproterozoic. *Geophysical Research Letters* **31** L05202

Lowell, R.P. & Keller, S. M. (2003). High-temperature seafloor hydrothermal circulation over geologic time and Archean banded iron formations. *Geophysical Research Letters* **30**(7) doi:10.1029/2002GL016536

Lovelock, J. E. & Whitfield, M. (1982). Life span of the biosphere. *Nature* **296** pp.561 - 563

Mills, B., Lenton, T. M., & Watson, A.J. (2014). Proterozoic oxygen rise linked to shifting balance between seafloor and terrestrial weathering *PNAS* **111** (25) pp. 9073 - 9078

O'Malley-James, J.T., Greaves, J.S., Raven, J.A. & Cockell, C.S. (2013). Swansong biospheres: refuges for life and novel microbial biospheres on terrestrial planets near the end of their habitable lifetimes. *International Journal of Astrobiology* **12** 99-112.

O'Malley-James, J.T., Greaves, J.S., Raven, J.A. & Cockell, C.S. (2014). Swansong biospheres II: The final signs of life on terrestrial planets near the end of their habitable lifetimes. *International Journal of Astrobiology,* **13** 3, 229 - 243

Omta, A.W., Dutkiewicz, S. & Follows, M.J. (2011). Dependence of the ocean atmosphere partitioning of carbon on temperature and alkalinity. *Global Biogeochemical Cycles*, **25** GB1003

O'Neill, C.O. and Lenardic, A. (2007). Geological consequences of supersized Earths. *Geophysical Research Letters* **34** L19204

Pierrehumbert, R. (2010). *Principles of Planetary Climate*. 1st ed. Cambridge: Cambridge University Press.

Rogers, L. (2015). Most 1.6 Earth-Radius Planets are not Rocky. *The Astrophysical Journal* **801** (41).

Rosing, M.T., Bird, D.K., Sleep, N.H. & Bjerrum, C.J. (2010). No climate paradox under the faint early Sun. *Nature* **464** pp. 744 - 747

Rothschild, L.J. (2007), In: Pudritz, R., Higgs, P. and Stone, J. (2007), *Planetary Systems and the Origins of Life*. Cambridge: Cambridge University Press

Royer, D.L, Berner, R.A., Montañez, I.P, Tabor, N.J., Beerling, D.J. (2004) $CO_2$ as a primary driver of Phanerozoic climate. *GSA Today* **14**(3) doi: 10.1130/1052-5173(2004)014

Rushby, A.J., Claire, M.W., Osborn, H. & Watson, A.J (2013). Habitable Zone Lifetimes of Exoplanets Around Main Sequence Stars. *Astrobiology,* **13**(9) 833 – 849

Rye, R., P. H. Kuo, & H. D. Holland (1995), Atmospheric carbon dioxide concentrations before 2.2 billion years ago. Nature 378, 603–605, doi:10.1038/378603a0.

Sackmann, I. J., Boothroyd, A. I., and Kraemer, K. E. (1993). Our sun. III. Present and future. *The Astrophysical Journal,* **418,** 457.

Schwartzman, D.W. & Volk, T. (1989). Biotic enhancement of weathering and the habitability of Earth. *Nature* **340** pp. 457 460

Seager, S., Kuchner, M., Hier-Majumder, C.A., Militzer, B. (2007). Mass-Radius Relationships for Solid Exoplanets *The Astrophysical Journal* **669** pp.1279 – 1297

Sheldon, N. D. (2006), Precambrian paleosols and atmospheric CO2 levels, Precambrian Research 147, 148–155, doi:10.1016/j.precamres.2006.02.004.

Shields, A.L., Ballard, S. & Asher Johnson, J. (2016). The habitability of planets oribiting M-dwarf stars. *Physics Reports* **663** pp. 1 - 38

Sleep, N. H. & Zahnle, K. (2001). Carbon dioxide cycling and implications for climate on ancient Earth. Journal of Geophysical Research 106 (E1) pp.1373 - 1399

Sotin, C., Jackson, J.M. & Seager, S. (2010). Terrestrial Planet Interiors in: Seager, S. (eds.) (2010) *Exoplanets* (1st edition). Tucson: Arizona University Press.

Swain, M. R., Vasisht, G. & Tinetti, G. (2008). The presence of methane in the atmosphere of an extrasolar planet. *Nature* **452** pp. 329 - 331

Szathmry, E. & Maynard-Smith, J. (1995) The Major Evolutionary Transitions. *Nature*. **374** pp. 227 - 232

Takai, K., Nakamura, K., Toki, T., Tsunogai, U., Miyazaki, M., Miyazaki, J.-I., Hirayama, H., Makagawa, S., Nunoura, T., & Horikoshi, K. (2008). Cell proliferation at 122 degrees C and isotopically heavy CH4 production by a hyperthermophilic methanogen under high-pressure cultivation. *Proceedings of the National Academy of Science* **105**(10) pp. 949 - 10,954.

Trenberth, K.E., Smith, L., Qian, T., Dai, A., & Fasullo, J. (2007). Estimates of the Global Water Budget and Its Annual Cycle Using Observational and Model Data. *Journal of Hydrometeorology* **8** pp.758 - 769

Valencia, D., OConnell, R.J. & Sasselov, D.D. (2007). Inevitability of Plate Tectonics on Super-Earths. *The Astrophysical Journal* **670** pp. 45 – 48

Walker, J.C.G., Hays, P.B., & Kasting, J.F. (1981). A negative feedback mechanism for the long-term stabilization of Earth's surface temperature. *Journal of Geophysical Research* **86** (10) pp. 9776 - 9782

Watson, A.J. (2008) Implications of an Anthropic Model of Evolution for Emergence of

Complex Life and Intelligence. *Astrobiology* **8**(1), pp. 1 - 12

West, A.J., Galy, A., & Bickle, M. (2005). Tectonic and climatic controls on silicate weathering. *Earth and Planetary Science Letters* **235** pp. 211 - 228

*Figure and Table Captions*

**Fig.1** : Model schematic illustrating carbon reservoirs and fluxes.

**Fig.2 :** Ensemble of sensitivity analyses **a)** Model output under differing values of κ (biotic enhancement factor, where 1 = present day weathering rates) compared to results from COPSE model (in dark green, with time-averaged values represented in grey). **b)** Modelled $p$CO$_2$ from this work(in bars) (dark line) as a function of time/solar constant, and comparison with empirical estimates of atmospheric carbon dioxide. Figure modified from Feulner (2012). The dashed line indicates pre-industrial $p$CO$_2$ (0.000280 bar)

**Fig.3** : **a)** Average surface temperature as a function of planet size and time, at 1 AU from a brightening Sun. Also displayed are insolation limits as a function of time. **b)** Atmospheric $p$CO$_2$, in pre-industrial units (RCO$_2$ = 280 ppm), as a function of planet size and time, at 1 AU from a brightening Sun. Also displayed are insolation limits as a function of time.

**Fig.4** : Contours of average surface temperature in K as a function of planet radius (in Earth-relative units) and incident flux (relative to present day Earth). Also displayed are current Earth (blue marker), and a several exoplanet candidates that fall within this parameter space.

**Fig.5** : Contours of atmospheric CO$_2$, in pre-industrial units (where 1 RCO$_2$ = 280 ppm) as a function of planet radius (in Earth-relative units) and incident flux (relative to present day Earth). Also displayed are the C3 and C4 photosynthesis limits.

**Table.1** : Definitions of terms used in this paper.

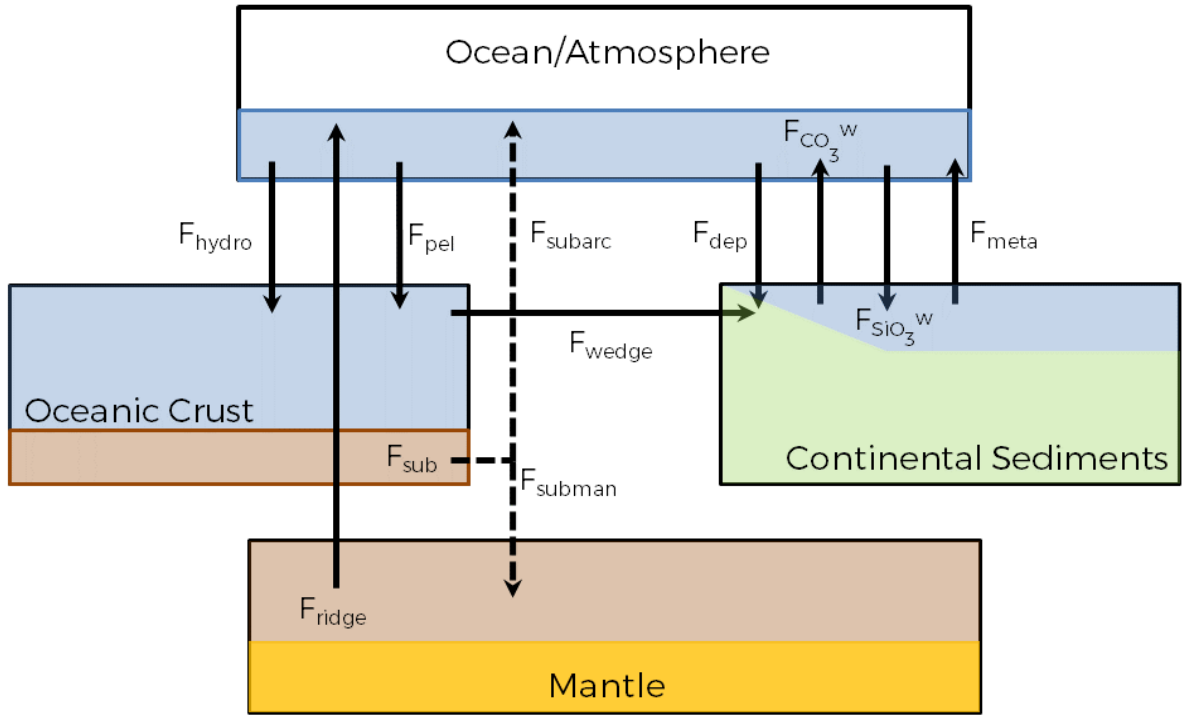

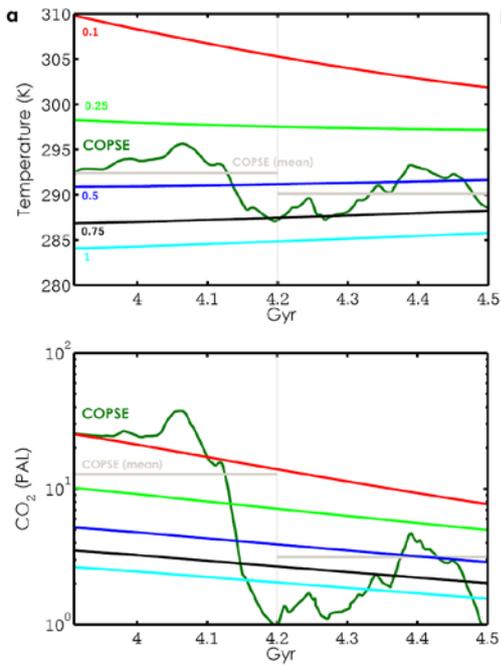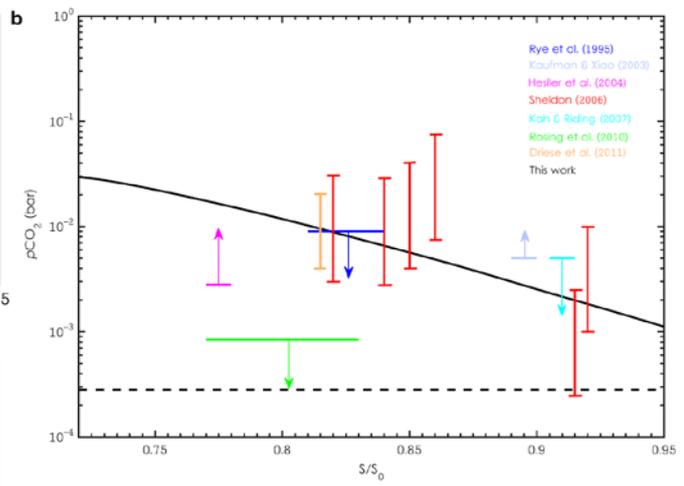

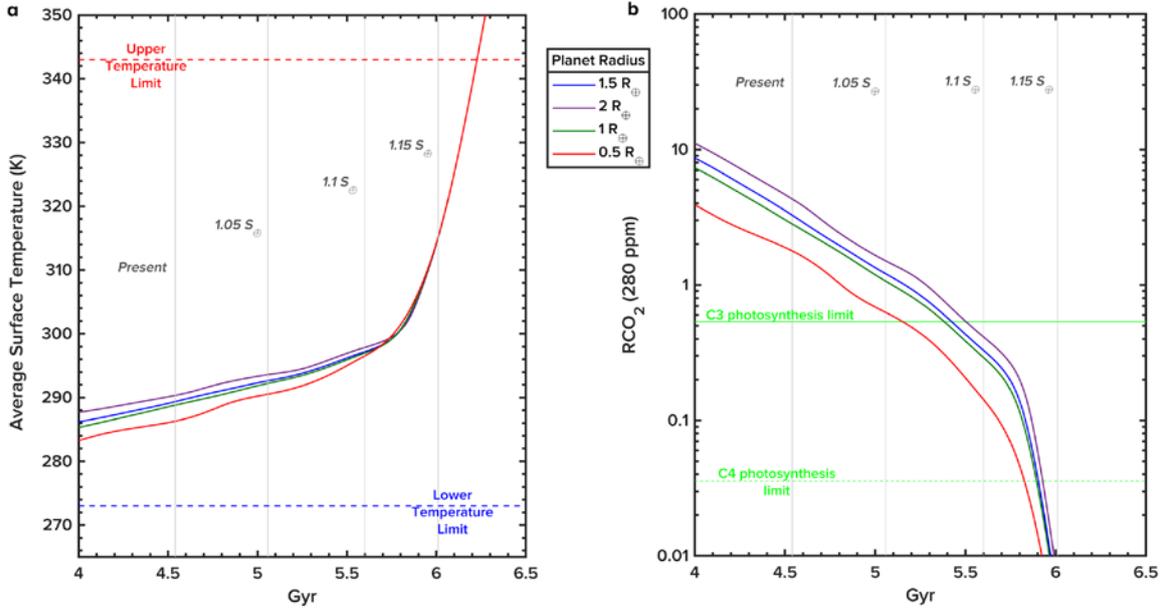

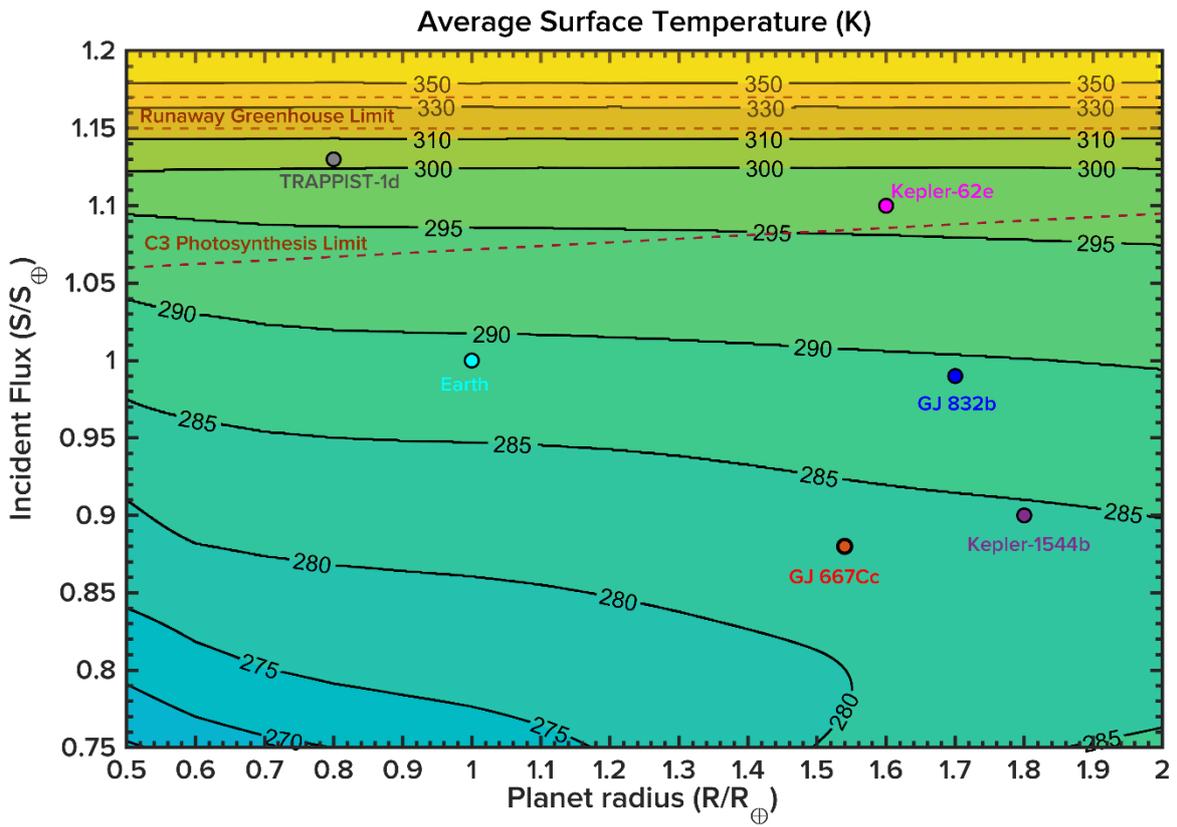

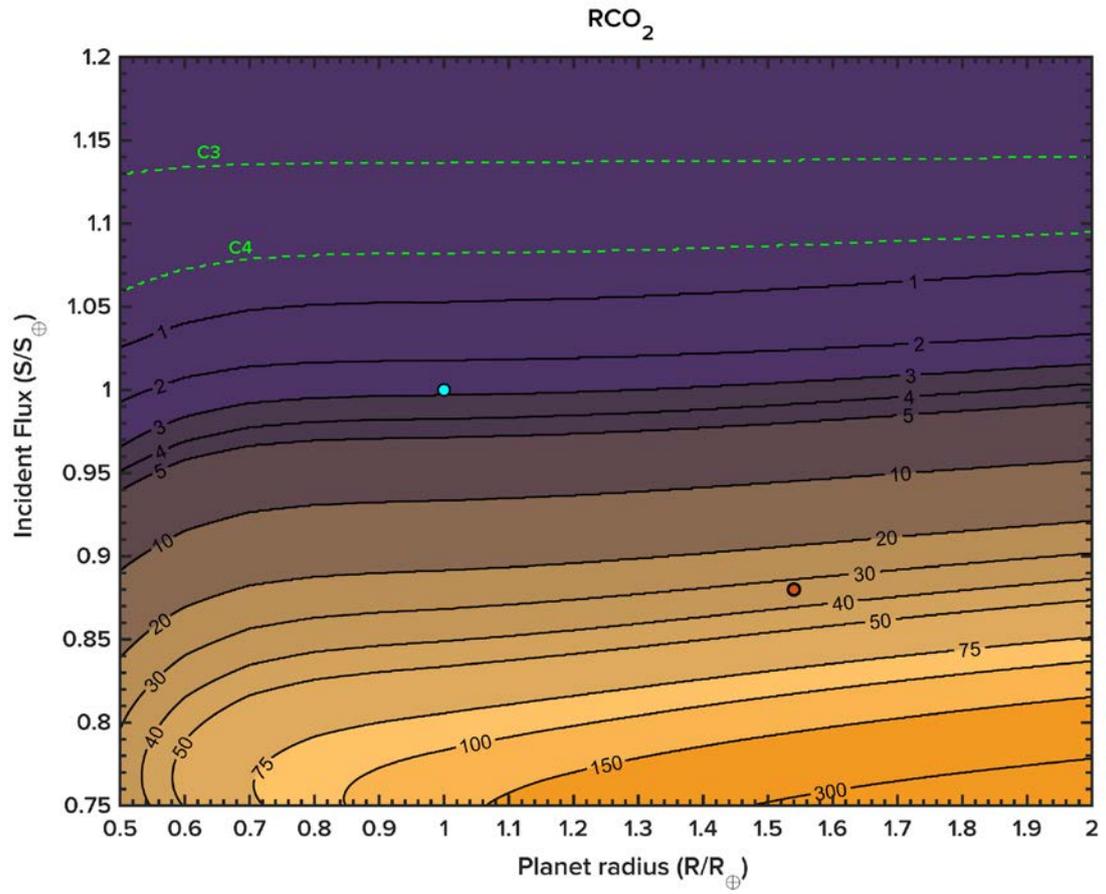

Table 1.

| Parameter | Definition | Value |
|---|---|---|
| $R_\oplus$ | Earth radius | $6.371 \times 10^6$ m |
| $M_\oplus$ | Earth mass | $5.98 \times 10^{24}$ kg |
| $S_\oplus$ | Present day solar irradiation (top of atmosphere) | 1367 W m$^{-2}$ |
| $F_{SiO_3 w_0}$ | Present day, steady state silicate weathering flux | $6.5 \times 10^{12}$ mols C yr$^{-1}$ |
| $Q_\oplus$ | Present day geothermal heat flux on Earth. | 0.0961 W m$^{-2}$ [1] |
| $\mu$ | Heat flow scaling parameter | -0.7 |
| $\beta$ | Controls the sensitivity of the terrestrial weathering to the partial pressure of atmospheric carbon dioxide | 0.3 |
| $B$ | Weathering calibration parameter | 13.7 |
| $R_E$ | Normalised rate of evaporation | $R/R_\oplus^2 \cdot (1 - C \cdot A_{ocean} / A_{planet})$ |
| $C$ | Constant that accounts for the increased rate of evaporation over the ocean. | $C = E_{land} / E_{ocean} \approx 0.176$ [2] |
| $\kappa$ | Bioamplification of terrestrial weathering i.e. the rate at which weathering proceeded without biology, relative to the present day. | 0.5 |
| $RCO_2$ | Units of preindustrial $p$CO$_2$ | 280 ppm |

[1] Davies & Davies, 2010
[2] Trenberth *et al.*, 2007